\documentstyle[12pt]{article}
\newcommand\be{\begin{equation}}
\newcommand\ee{\end{equation}}
\newcommand\bea{\begin{eqnarray}}
\newcommand\eea{\end{eqnarray}}
\newcommand\ket[1]{|#1\rangle}

\newcommand\braket[2]{\langle #1|#2\rangle}
\newcommand{\fatalpha}{{\bf \alpha \kern -0.44em \alpha}}
\newcommand{\fatsigma}{{\bf \sigma \kern -0.54em \sigma}}
\newcommand{\tpchi}{{\bf \chi \kern -0.35em \chi}}
\newcommand{\llambda}{{\bf \lambda \kern -0.45em \lambda}}

\renewcommand{\theequation}{\arabic{equation}}
\renewcommand{\theequation}{\thesection-\arabic{equation}}
\bibliography{plain}
\title{\bf Calculating two-point resistances in distance-regular resistor networks}\vspace{20mm}
\author{ M. A. Jafarizadeh$^{a,b,c}$
 \thanks{E-mail:jafarizadeh@tabrizu.ac.ir},
 R. Sufiani$^{a,c}$,
 S. Jafarizadeh$^{d}$
\thanks{E-mail:sofiani@tabrizu.ac.ir}
\\ $^a${\small Department of Theoretical Physics and Astrophysics,
Tabriz University, Tabriz 51664, Iran.} \\ $^b${\small Institute
for Studies in Theoretical Physics and Mathematics, Tehran
19395-1795, Iran.} \\ $^c${\small Research Institute for
Fundamental Sciences, Tabriz 51664, Iran.}\\ $^d${\small
Department of Electrical and computer engineering, Tabriz
University, Tabriz 51664, Iran.}} \pagebreak

\vspace{20mm}
\begin{document}
\maketitle \vspace{15mm}
\newpage
\begin{abstract}
An algorithm for the calculation of the resistance between two
arbitrary nodes in an arbitrary distance-regular resistor network is
provided, where the calculation is based on stratification
introduced in \cite{js} and Stieltjes transform of the spectral
distribution (Stieltjes function) associated with the network. It is
shown that the resistances between a node $\alpha$ and all nodes
$\beta$ belonging to the same stratum with respect to the $\alpha$
($R_{\alpha\beta^{(i)}}$, $\beta$ belonging to the $i$-th stratum
with respect to the $\alpha$) are the same. Also, the analytical
formulas for two-point resistances $R_{{\alpha\beta^{(i)}}},
i=1,2,3$ are given in terms of the the size of the network and
corresponding intersection numbers. In particular, the two-point
resistances in a strongly regular network are given in terms of the
its parameters ($v,\kappa,\lambda,\mu$). Moreover, the lower and
upper bounds for two-point resistances in strongly regular networks
are discussed.


{\bf PACs Index: 01.55.+b, 02.10.Yn }
\end{abstract}

\vspace{70mm}
\newpage
\section{Introduction}
A classic problem in electric circuit theory studied by numerous
authors over many years is the computation of the resistance between
two nodes in a resistor network (see, e.g., \cite{Cserti}). Besides
being a central problem in electric circuit theory, the computation
of resistances is also relevant to a wide range of problems ranging
from random walks (see \cite{2}), the theory of harmonic functions
\cite{4}, to lattice Green's functions \cite{6,6a,6b,6c,6d}. The
connection with these problems originates from the fact that
electrical potentials on a grid are governed by the same difference
equations as those occurring in the other problems. For this reason,
the resistance problem is often studied from the point of view of
solving the difference equations, which is most conveniently carried
out for infinite networks. In the case of Green's function approach,
for example, past efforts \cite{Cserti}, \cite{7} have been focused
mainly on infinite lattices. Little attention has been paid to
finite networks, even though the latter are those occurring in real
life. In this paper, we take up this problem and present a general
formulation for computing two-point resistances in finite networks.
Particularly, we show that known results for infinite networks are
recovered by taking the infinite-size limit.

The study of electric networks was formulated by Kirchhoff \cite{8}
more than 150 years ago as an instance of a linear analysis. Our
starting point is along the same line by considering the Laplacian
matrix associated with a network. The Laplacian is a matrix whose
off-diagonal entries are the conductances connecting pairs of nodes.
Just as in graph theory where everything about a graph is described
by its adjacency matrix (whose elements is 1 if two vertices are
connected and 0 otherwise), everything about an electric network is
described by its Laplacian. The author of \cite{Wu}, has been
derived an expression for the two-point resistance between two
arbitrary nodes $\alpha$ and $\beta$ of a regular network in terms
of the matrix entries $L^{-1}_{\alpha\alpha}$, $L^{-1}_{\beta\beta}$
and $L^{-1}_{\alpha\beta}$, where $L^{-1}$ is the pseudo inverse of
the Laplacian matrix. Here in this work, based on stratification
introduced in \cite{js} and spectral analysis method, we introduce
an procedure for calculating two-point resistances in
distance-regular resistor networks in terms of the Stieltjes
function $G_{\mu}(x)$  associated with the adjacency matrix of the
network and its derivatives. Although, we discuss the case of
distance-regular networks, but also the method can be used for any
arbitrary regular network. It should be noticed that, in this way,
the two-point resistances are calculated straightforwardly without
any need to know the spectrum of the network. Also, it is shown that
the resistances between a node $\alpha$ and all nodes $\beta$
belonging to the same stratum with respect to the $\alpha$
($R_{\alpha\beta^{(i)}}$, $\beta$ belonging to the $i$-th stratum
with respect to the $\alpha$) are the same. We give  the analytical
formulas for two-point resistances $R_{{\alpha\beta^{(i)}}},i=1,2,3$
in terms of the network's characteristics such as the size of the
network and its intersection array. In particular, the two-point
resistances in a strongly regular network  are given in terms of the
network's parameters  ($v,\kappa,\lambda,\mu$). Moreover, we discuss
the lower and upper bounds for two-point resistances in strongly
regular networks. From the fact that, the two-point resistances on a
network depend on the corresponding Stieltjes function $G_{\mu}(x)$
and that $G_{\mu}(x)$ is written as a continued fraction, the
two-point resistances on an infinite-size network can be
approximated with those of the corresponding finite-size networks.

The organization of the paper is as follows. In section 2, we give
some preliminaries such as association schemes, distance-regular
networks, stratification of these networks and Stieltjes function
associated with the network. In section $3$, two-point resistances
in distance-regular networks are given in terms of the Stieltjes
function and its derivatives. Also, the resistances
$R_{\alpha\beta^{(i)}}$, for $i=1,2,3$ are given in terms of the
network's intersection array. In particular, two-point resistances
in a strongly regular network are given in terms of the network's
parameters, also lower and upper bounds for the two-point
resistances in these networks are discussed. Section $4$ is devoted
to calculating two-point resistances $R_{\alpha\beta^{(i)}}$ for
$i=1,2,3$ in some important examples of distance-regular networks,
such as complete network, strongly regular networks
(distance-regular networks with diameter $2$), e.g. Petersen and
normal subgroup scheme networks \cite{js}, $d$-cube ($d$ dimensional
hypercube) and Johnson networks. The paper is ended with a brief
conclusion and an appendix containing   a table for  two-point
resistances $R_{\alpha\beta^{(i)}}$, $i=1,2,3$ of some important
distance-regular resistor networks with size less than $70$.
\section{Preliminaries}
In this section we give some preliminaries such as definitions
related to association schemes, corresponding stratification,
distance-regular networks and Stieltjes function associated with the
network.
\subsection{Association schemes}
First we recall the definition of association schemes. The reader
is referred to Ref.\cite{Ass.sch.}, for further information on
association
schemes.\\
\textbf{Definition 2.1} (Symmetric association schemes). Let $V$
be a set of vertices, and let $R_i(i = 0, 1,..., d)$ be nonempty
relations on $V$ (i.e., subset of $V\times V$). Let the following
conditions (1), (2), (3) and (4) be satisfied. Then, the relations
$\{R_i\}_{0\leq i\leq d}$
on $V\times V$ satisfying the following conditions\\
$(1)\;\ \{R_i\}_{0\leq i\leq d}$ is a partition of $V\times V$\\
$(2)\;\ R_0=\{(\alpha, \alpha) : \alpha\in V \}$\\
$(3)\;\ R_i=R_i^t$ for $0\leq i\leq d$, where
$R_i^t=\{(\beta,\alpha) :(\alpha, \beta)\in R_i\} $\\
$(4)$ For $(\alpha, \beta)\in R_k$, the number  $p^k_{i,j}=\mid
\{\gamma\in X : (\alpha, \beta)\in R_i \;\ and \;\
(\gamma,\beta)\in R_j\}\mid$ does not depend on $(\alpha, \beta)$
but only on $i,j$ and $k$,\\ define a symmetric association scheme
of class $d$ on $V$ which is denoted by $Y=(V,\{R_i\}_{0\leq i\leq
d})$. Furthermore, if we have $p^k_{ij}=p^k_{ji}$ for all
$i,j,k=0,2,...,d$, then $Y$ is called commutative.

The intersection number $p_{ij}^k$ can be interpreted as the
number of vertices which have relation $i$ and $j$ with vertices
$\alpha$ and $\beta$, respectively provided that
$(\alpha,\beta)\in R_k$, and it is the same for all element of
relation $R_k$. For all integers $i$ ($0 \leq i \leq d$), set
$k_i=p_{ii}^{0}$ and note that $k_i\neq 0$, since $R_i$ is
non-empty. We refer to $k_i$ as the $i$-th valency of $Y$.

Let $Y=(X,\{R_i\}_{0\leq i\leq d})$ be a commutative symmetric
association scheme of class $d$, then the matrices
$A_0,A_1,...,A_d$ defined by
\begin{equation}\label{adj.}
\bigl(A_{i})_{\alpha, \beta}\;=\;\cases{1 & if $\;(\alpha,
\beta)\in R_i$ \cr 0 & otherwise\cr},
\end{equation}
are adjacency matrices of $Y$ such that
\begin{equation}\label{ss}
A_iA_j=\sum_{k=0}^{d}p_{ij}^kA_{k}.
\end{equation}
From (\ref{ss}), it is seen that the adjacency matrices $A_0, A_1,
..., A_d$ form a basis for a commutative algebra \textsf{A} known
as the Bose-Mesner algebra of $Y$.\\
\textbf{Definition 2.2} ($P$-polynomial property)
$Y=(X,{\{R_i\}}_{0\leq i\leq d})$ is said to be $P$-polynomial
(with respect to the ordering $R_0,..., R_d$ of the associate
classes) whenever for all integers $h, i, j$ ($0\leq h, i, j\leq
d$),
$$p^h_{ij}=0\;\;\ \mbox{if one of}\;\;\ h, i, j \;\;\ \mbox{is greater than the sum of the other
two},$$
\begin{equation}
p^h_{ij}\neq 0\;\;\ \mbox{if one of}\;\;\ h, i, j\;\;\
\mbox{equals the sum of the other two}.
\end{equation}
It is shown in \cite{bannai} that in the case of P-polynomial
schemes, $A_i=P_i(A)$ $(0\leq i\leq d)$, where  $P_i$ is a
polynomial with real coefficients and degree exactly $i$. In
particular, $A$ multiplicatively generates the Bose-Mesner
algebra.

Finally the underlying graph of an association scheme
$\Gamma=(V,R_1)$ is an undirected connected graph, where the set
$V$ and $R_1$ consist of its vertices and edges, respectively.
Obviously replacing $R_1$ with one of other relation such as
$R_i$, for  $i\neq 0,1$ will also give us an underlying graph
$\Gamma=(V,R_i)$ (not necessarily a connected graph ) with the
same set of vertices but a new set of edges $R_i$.
\subsection{Stratifications}
For a given vertex $\alpha\in V$, let $R_i(\alpha):=\{\beta\in V:
(\alpha, \beta)\in R_i\}$  denote the set of all vertices having
the relation $R_i$ with $\alpha$. Then, the vertex set $V$ can be
written as disjoint union of $R_i(\alpha)$ for $i=0,1,2,...,d$,
i.e.,
 \begin{equation}\label{asso1}
 V=\bigcup_{i=0}^{d}R_{i}(\alpha).
 \end{equation}
We fix a point $o\in V$ as an origin of a distance regular graph
(the underlying graph of an association scheme), called reference
vertex. Then, the relation (\ref{asso1}) stratifies  the
underlying graph into a disjoint union of associate classes
$R_{i}(o)$. With each associate class $R_{i}(o)$ we associate a
unit vector in $l^2(V)$ defined by
\begin{equation}\label{unitv}
\ket{\phi_{i}}=\frac{1}{\sqrt{\kappa_i}}\sum_{\alpha\in
R_{i}(o)}\ket{\alpha},
\end{equation}
where, $\ket{\alpha}$ denotes the eigenket of $\alpha$-th vertex
at the associate class $R_{i}(o)$ and $\kappa_i=|R_{i}(o)|$ is
called the $i$-th valency of the graph (i.e.,
$\kappa_i=p^0_{i,i}$).
  The closed subspace of $l^2(V)$ spanned by
$\{\ket{\phi_{i}}\}$ is denoted by $\Lambda(G)$. Since
$\{\ket{\phi_{i}}\}$ becomes a complete orthonormal basis of
$\Lambda(G)$, we often write
\begin{equation}
\Lambda(G)=\sum_{i}\oplus \textbf{C}\ket{\phi_{i}}.
\end{equation}
Let $A_i$ be the adjacency matrix of a graph $\Gamma=(V,R)$ for
reference state $\ket{\phi_0}$ ($\ket{\phi_0}=\ket{o}$, with $o\in
V$ as reference vertex), we have
\begin{equation}\label{Foc1}
A_i\ket{\phi_0}=\sum_{\beta\in R_{i}(o)}\ket{\beta}.
\end{equation}
Then by using unit vectors $\ket{\phi_{i}}$ (\ref{unitv}) and
(\ref{Foc1}), we have
\begin{equation}\label{Foc2}
A_i\ket{\phi_0}=\sqrt{\kappa_i}\ket{\phi_i}.
\end{equation}
\subsection{Distance-regular graphs}
Here in this section we consider some set of important
graphs called distance regular graphs, where the relations are
based on distance function defined as follows: A finite sequence
$\alpha_0, \alpha_1, . . . , \alpha_n \in V$ is called a walk of
length $n$ (or of $n$ steps) if $\alpha_{k-1}\sim \alpha_k$ for
all $k=1, 2, . . . , n$. For $\alpha\neq \beta$ let
$\partial(\alpha, \beta)$  be the length of the shortest walk
connecting $\alpha$ and $\beta$, therefore $\partial(\alpha,
\beta)$ gives the distance between vertices $\alpha$ and $\beta$
hence it is  called the distance function and we have
$\partial(\alpha, \alpha)=0$ for all $\alpha\in V$ and
$\partial(\alpha, \beta)=1$ if and only if $\alpha\sim \beta$.
Therefore, the distance regular graphs become  metric spaces with
the distance function $\partial$.

An undirected connected graph $\Gamma=(V,R_1)$ is called distance
regular graph if it is the underlying graph of a $P$-polynomial
association scheme with relations defined  as:  $(\alpha,
\beta)\in R_i$ if and only if $\partial(\alpha, \beta)=i$, for
$i=0,1,...,d$, where $d:=$max$\{\partial(\alpha, \beta): \alpha,
\beta\in V \}$ is called the diameter of the graph. Usually in
distance regular graphs, the relations $R_i$ are denoted by
$\Gamma_i$.

 Now, in any connected graph, for every
$\beta\in R_i(\alpha)$ we have
\begin{equation}
R_1(\beta)\subseteq R_{i-1}(\alpha)\cup R_i(\alpha)\cup
R_{i+1}(\alpha).
\end{equation}
Hence in a distance regular graph, $p_{j1}^i=0 $ (for $i\neq 0$,
$j$ is not $\{i-1, i, i+1 \}$). The intersection numbers  of the
graph are defined as
\begin{equation}\label{abc}
 a_i=p_{i1}^i, \;\;\;\  b_i=p_{i+1,1}^i, \;\;\;\
 c_i=p_{i-1,1}^i\;\ .
\end{equation}
The intersection numbers (\ref{abc}) and the valencies $\kappa_i$
satisfy the following obvious conditions
$$a_i+b_i+c_i=\kappa,\;\;\ \kappa_{i-1}b_{i-1}=\kappa_ic_i ,\;\;\
i=1,...,d,$$
\begin{equation}\label{intersec}
\kappa_0=c_1=1,\;\;\;\ b_0=\kappa_1=\kappa, \;\;\;\ (c_0=b_d=0).
\end{equation}
Thus all parameters of the graph can be obtained from the
intersection array $\{b_0,...,b_{d-1};c_1,...,c_d\}$.

It should be noticed that, for distance-regular graphs, the unit
vectors $\ket{\phi_i}$ for $i=0,1,...,d$ defined as in
(\ref{unitv}), satisfy the following three-term recursion
relations
\begin{equation}\label{trt}
A\ket{\phi_i}=\beta_{i+1}\ket{\phi_{i+1}}+\alpha_i\ket{\phi_i}+\beta_{i}\ket{\phi_{i-1}},
\end{equation}
where, the coefficients $\alpha_i$ and $\beta_i$ are defined
as
\begin{equation}\label{omegal}
\alpha_k=\kappa-b_{k}-c_{k},\;\;\;\;\
\omega_k\equiv\beta^2_k=b_{k-1}c_{k},\;\;\ k=1,...,d,
\end{equation}
i.e., in the basis of unit vectors $\{\ket{\phi_i},i=0,1,...,d\}$,
the adjacency matrix $A$ is projected to the following symmetric
tridiagonal form:
\begin{equation}\label{trid}
A=\left(
\begin{array}{cccccc}
 \alpha_0 & \beta_1 & 0 & ... &...&0 \\
      \beta_1 & \alpha_1 & \beta_2 & 0 &...&0 \\
      0 & \beta_2 & \alpha_3 & \beta_3 & \ddots&\vdots \\
     \vdots & \ddots &\ddots& \ddots &\ddots &0\\
     0 & \ldots  &0 &\beta_{d-1} & \alpha_{d-1} &\beta_{d}\\
          0&... & 0 &0 & \beta_{d} & \alpha_{d}\\
\end{array}
\right).
\end{equation}
In Ref. \cite{jss}, it has be shown that, the coefficients
$\alpha_i$ and $\beta_i$ can be also obtained easily by using the
Lanczos iteration algorithm. Hereafter, we will refer to the
parameters $\alpha_i$ and $\omega_i$ in (\ref{omegal}) as QD
(Quantum decomposition) parameters.

One should notice that, in distance regular graphs stratification
is reference state independent, namely we can choose every vertex
as a reference state, while the stratification of more general
graphs may be reference dependent.
\subsection{Stieltjes function associated with the network }
It is well known that, for any pair $(A,\ket{\phi_0})$ of a matrix
$A$ and a vector $\ket{\phi_0}$, it can be assigned a measure
$\mu$ as follows
\begin{equation}\label{sp1}
\mu(x)=\braket{ \phi_0}{E(x)|\phi_0},
\end{equation}
 where
$E(x)=\sum_i|u_i\rangle\langle u_i|$ is the operator of projection
onto the eigenspace of $A$ corresponding to eigenvalue $x$, i.e.,
\begin{equation}
A=\int x E(x)dx.
\end{equation}
It is easy to see that, for any polynomial $P(A)$ we have
\begin{equation}\label{sp2}
P(A)=\int P(x)E(x)dx,
\end{equation}
where for discrete spectrum the above integrals are replaced by
summation. Therefore, using the relations (\ref{sp1}) and
(\ref{sp2}), the expectation value of powers of adjacency matrix
$A$ over starting site $\ket{\phi_0}$ can be written as
\begin{equation}\label{v2}
\braket{\phi_{0}}{A^m|\phi_0}=\int_{R}x^m\mu(dx), \;\;\;\;\
m=0,1,2,....
\end{equation}
The existence of a spectral distribution satisfying (\ref{v2}) is
a consequence of Hamburger's theorem, see e.g., Shohat and
Tamarkin [\cite{st}, Theorem 1.2].

Obviously relation (\ref{v2}) implies an isomorphism from the
Hilbert space of the stratification onto the closed linear span of
the orthogonal polynomials with respect to the measure $\mu$. It
is well known that \cite{jss} for distance-regular graphs we have
\begin{equation}\label{xx}
\ket{\phi_i}=P_i(A)\ket{\phi_0},
\end{equation}
where $P_i$ is a polynomial with real coefficients and degree $i$.
Now, substituting (\ref{xx}) in (\ref{trt}), we get three term
recursion relations between polynomials $P_j(A)$, which leads to
the following  three term recursion between polynomials $P_j(x)$
\begin{equation}\label{trt0}
xP_{k}(x)=\beta_{k+1}P_{k+1}(x)+\alpha_kP_{k}(x)+\beta_kP_{k-1}(x)
\end{equation}
for $k=0,...,d-1$, with $P_0(x)=1$.\\
Multiplying by $\beta_1...\beta_k$ we obtain
\begin{equation}
\beta_1...\beta_kxP_{k}(x)=\beta_1...\beta_{k+1}P_{k+1}(x)+\alpha_k\beta_1...\beta_kP_{k}(x)+\beta_k^2.\beta_1...\beta_{k-1}P_{k-1}(x).
\end{equation}
By rescaling $P_k$ as $Q_k=\beta_1...\beta_kP_k$, the spectral
distribution $\mu$ under question is characterized by the property
of orthonormal polynomials $\{Q_k\}$ defined recurrently by
$$ Q_0(x)=1, \;\;\;\;\;\
Q_1(x)=x,$$
\begin{equation}\label{op}
xQ_k(x)=Q_{k+1}(x)+\alpha_{k}Q_k(x)+\beta_k^2Q_{k-1}(x),\;\;\
k\geq 1.
\end{equation}

If such a spectral distribution is unique, the spectral
distribution $\mu$ is determined by the identity
\begin{equation}\label{sti}
G_{\mu}(x)=\int_{R}\frac{\mu(dy)}{x-y}=\frac{1}{z-\alpha_0-\frac{\beta_1^2}{z-\alpha_1-\frac{\beta_2^2}
{z-\alpha_2-\frac{\beta_3^2}{z-\alpha_3-\cdots}}}}=\frac{Q_{d-1}^{(1)}(x)}{Q_{d}(x)}=\sum_{l=0}^{d-1}
\frac{A_l}{x-x_l},
\end{equation}
where, $x_l$ are the roots  of polynomial $Q_{d}(x)$. $G_{\mu}(z)$
is called the Stieltjes/Hilbert transform of spectral distribution
$\mu$ or Stieltjes function  and polynomials $\{Q_{k}^{(1)}\}$ are
defined recurrently as
$$Q_{0}^{(1)}(x)=1, \;\;\;\;\;\
    Q_{1}^{(1)}(x)=x-\alpha_1,$$
\begin{equation}\label{oq}
xQ_{k}^{(1)}(x)=Q_{k+1}^{(1)}(x)+\alpha_{k+1}Q_{k}^{(1)}(x)+\beta_{k+1}^2Q_{k-1}^{(1)}(x),\;\;\
k\geq 1,
\end{equation}
respectively. The coefficients $A_l$ appearing in (\ref{sti}) are
calculated as
\begin{equation}\label{Gauss}
A_l=\lim_{z\rightarrow x_l}(z-x_l)G_{\mu}(z).
\end{equation}
(for more details see Refs.\cite{st, obh,tsc,obah}.)
\section{Two-point resistances in regular resistor networks}
A classic problem in electric circuit theory studied by numerous
authors over many years, is the computation of the resistance
between two nodes in a resistor network (see, e.g., \cite{Cserti}).
The results obtained   in this section show that, there is a close
connection between the techniques introduced in section $2$ such as
Hilbert space of the stratification and the Stieltjes function and
electrical concept of resistance between two arbitrary nodes of
regular networks and these techniques can be employed for
calculating two-point resistances.

For a given regular graph $\Gamma$ with $n$ vertices and adjacency
matrix $A$, let $r_{ij}=r_{ji}$ be the resistance of the resistor
connecting vertices $i$ and $j$. Hence, the conductance is
$c_{ij}=r^{-1}_{ij}=c_{ji}$ so that $c_{ij}=0$ if there is no
resistor connecting $i$ and $j$. Denote the electric potential at
the $i$-th vertex by $V_i$ and the net current flowing into the
network at the $i$-th vertex by $I_i$ (which is zero if the $i$-th
vertex is not connected to the external world). Since there exist
no sinks or sources of current including the external world, we
have the constraint $\sum_{i=1}^nI_i=0$. The Kirchhoff law states
\begin{equation}\label{resistor}
\sum_{j=1,j\neq i}^nc_{ij}(V_i-V_j)=I_i,\;\;\  i=1,2,...,n.
\end{equation}
Explicitly, Eq.(\ref{resistor}) reads
\begin{equation}\label{resistor1}
L\vec{V}=\vec{I},
\end{equation}
where, $\vec{V}$ and $\vec{I}$ are $n$-vectors whose components
are $V_i$ and $I_i$, respectively and
\begin{equation}\label{laplas}
L=\sum_{i}c_i|i\rangle\langle i|-\sum_{i,j}c_{ij}|i\rangle\langle
j|
\end{equation}
is the Laplacian of the graph $\Gamma$ with
\begin{equation}
c_i\equiv \sum_{j=1,j\neq i}^nc_{ij},
\end{equation}
for each vertex $\alpha$. Hereafter, we will assume that all
nonzero resistances are equal to $r$, then the off-diagonal
elements of $-L$ are precisely those of $\frac{1}{r}A$, i.e.,
\begin{equation}\label{laplas1}
L=\frac{1}{r}(\kappa I-A),
\end{equation}
with $\kappa=deg(\alpha)$, for each vertex $\alpha$. It should be
noticed that, $L$ has eigenvector $(1,1,...,1)^t$ with eigenvalue
$0$. Therefore, $L$ is not invertible and so we define the
psudo-inverse of $L$ as
\begin{equation}\label{inv.laplas}
L^{-1}=\sum_{i,\lambda_i\neq0} {\lambda}^{-1}_iE_i,
\end{equation}
where, $E_i$ is the operator of projection onto the eigenspace of
$L^{-1}$ corresponding to eigenvalue $\lambda_i$. Following the
result of \cite{Wu} and that $L^{-1}$ is a real matrix, the
resistance between vertices $\alpha$ and $\beta$ is given by
\begin{equation}\label{eq.res.}
R_{\alpha\beta}=\langle \alpha|L^{-1}|\alpha\rangle-2\langle
\alpha|L^{-1}|\beta\rangle+\langle \beta|L^{-1}|\beta\rangle.
\end{equation}

In this paper, we will consider distance-regular graphs as
resistor networks. Then, the diagonal entries of $L^{-1}$ are
independent of the vertex, i.e.,
$L^{-1}_{\alpha\alpha}=L^{-1}_{\beta\beta}$ for all
$\alpha,\beta\in V$. Therefore, from the relation (\ref{eq.res.}),
we can obtain the two-point resistance between two arbitrary nodes
$\alpha$ and $\beta$ as follows
 is written as
\begin{equation}\label{eq.res.dist.}
R_{\alpha\beta}=2(L^{-1}_{\alpha\alpha}-L^{-1}_{\alpha\beta}).
\end{equation}

Now, let $\alpha$ and $\beta$ be two arbitrary nodes of the network
such that $\beta$ belongs to the $l$-th stratum with respect to
$\alpha$, i.e., $\beta\in R_l(\alpha)$ (we choose one of the nodes,
here $\alpha$, as reference node). Then, for calculating the matrix
entries $L^{-1}_{\alpha\alpha}$ and $L^{-1}_{\beta\alpha}$ in
(\ref{eq.res.}), we use the Stieltjes function to obtain
\begin{equation}\label{l1}
L^{-1}_{\alpha\alpha}=r\langle \alpha|\frac{1}{\kappa
I-A}|\alpha\rangle=r\int_{R-\{\kappa\}}\frac{d\mu(x)}{\kappa-x}=r\sum_{i,i\neq0}^{d-1}\frac{A_i}{\kappa-x_i}=r\lim_{y\rightarrow
\kappa}(G_{\mu}(y)-\frac{A_0}{y-\kappa})
\end{equation}
and
$$L^{-1}_{\beta\alpha}=r\langle \beta|\frac{1}{\kappa
I-A}|\alpha\rangle=\frac{r}{\sqrt{\kappa_l}}\langle
\phi_l|\frac{1}{\kappa
I-A}|\alpha\rangle=\frac{r}{\sqrt{\kappa_l}}\langle
\alpha|\frac{P_l(A)}{\kappa I-A}|\alpha\rangle$$
\begin{equation}\label{l2}
=\frac{r}{\sqrt{\kappa_l}}\int_{R-\{\kappa\}}\frac{d\mu(x)}{\kappa-x}P_l(x)=\frac{r}{\sqrt{\kappa_l}}\sum_{i,i\neq0}\frac{A_iP_l(x_i)}{\kappa-x_i},
\end{equation}
where, we have considered $x_0=\kappa$ ($\kappa$ is the eigenvalue
corresponding to the idempotent $E_0$). Then, by using
(\ref{eq.res.dist.}), the two-point resistance
$R_{\alpha\beta^{(l)}}$ in the network is given by
\begin{equation}\label{l3}
R_{\alpha\beta^{(l)}}=\frac{2r}{\sqrt{\kappa_l}}(\sqrt{\kappa_l}\lim_{y\rightarrow
\kappa}(G_{\mu}(y)-\frac{A_0}{y-\kappa})-\sum_{i,i\neq0}\frac{A_iP_l(x_i)}{\kappa-x_i}).
\end{equation}
For evaluating the term
$\sum_{i,i\neq0}\frac{A_iP_l(x_i)}{\kappa-x_i}$ in (\ref{l3}), we
need to calculate
\begin{equation}\label{eqq}
I_m:=\sum_{i,i\neq0}\frac{A_ix^m_i}{\kappa-x_i},\;\;\
\mbox{for}\;\;\ m=0,1,...,l.
\end{equation}
To do so, we write the term (\ref{eqq}) as
$$I_m=\sum_{i,i\neq0}\frac{A_ix^m_i}{\kappa-x_i}=\sum_{i,i\neq0}\frac{A_i((x_i-\kappa)^m-\sum_{l=1}^{m}(-1)^lC^m_l\kappa^lx^{m-l}_i)}{\kappa-x_i}$$
\begin{equation}\label{eqq2}
=-\sum_{i,i\neq0}A_i(x_i-\kappa)^{m-1}-\sum_{l=1}^{m}(-1)^lC^m_l\kappa^l\sum_{i,i\neq0}\frac{A_ix^{m-l}_i}{\kappa-x_i},
\end{equation}
that is, we have
\begin{equation}\label{eqq3}
I_m=-\sum_{l=0}^{m-1}(-1)^lC^{m-1}_l\kappa^l\sum_{i,i\neq0}A_ix^{m-l-1}_i-\sum_{l=1}^{m}(-1)^lC^{m}_l\kappa^l
I_{m-l}.
\end{equation}
Therefore, $I_m$ can be calculated recursively, if we able to
calculate the term $\sum_{i,i\neq0}A_ix^{m-l-1}_i$ for
$l=0,1,...,m-1$ appearing in (\ref{eqq2}). For example, for $m=1$,
we obtain
\begin{equation}\label{m=1}
I_1=\sum_{i,i\neq0}\frac{A_ix_i}{\kappa-x_i}=-\sum_{i,i\neq0}A_i+\kappa
I_0=-1+A_0+\kappa\sum_{i,i\neq0}\frac{A_i}{\kappa-x_i}.
\end{equation}
In order to evaluation of the sum $\sum_{i,i\neq0}A_ix^{k}_i$, we
rescale the roots $x_i$ as $\xi x_i$, where $\xi$ is some nonzero
constant. Then, we will have
\begin{equation}\label{st1}
\frac{1}{\xi}G_{\mu}(x/\xi)=\sum_{i}\frac{A_i}{x-\xi
x_i}+\frac{A_0}{x-\xi x_0}.
\end{equation}
Now, we take the $m$-th derivative of (\ref{st1}) to obtain
\begin{equation}\label{st11}
\frac{\partial^m}{\partial\xi^m}(\frac{1}{\xi}G_{\mu}(x/\xi))=m!(\sum_{i,i\neq0}\frac{A_ix^{m}_i}{(x-\xi
x_i)^{m+1}}+\frac{A_0x^{m}_0}{(x-\xi x_0)^{m+1}}),
\end{equation}
where, at the limit of the large $x$, one can obtain the following
simple form
\begin{equation}\label{st111}
\lim_{x\rightarrow\infty}\frac{\partial^m}{\partial\xi^m}(\frac{1}{\xi}G_{\mu}(x/\xi))=m!(\frac{\sum_{i,i\neq0}A_ix^{m}_i+A_0x^{m}_0}{x^{m+1}})
\end{equation}
Therefore, we obtain
\begin{equation}\label{st1111}
\sum_{i,i\neq0}A_ix^{m}_i=\frac{1}{m!}\lim_{x\rightarrow\infty}[x^{m+1}\frac{\partial^m}{\partial\xi^m}(\frac{1}{\xi}G_{\mu}(x/\xi))]-A_0x^{m}_0.
\end{equation}
\subsection{Two-point resistances up to the third stratum}
In this subsection we give analytical formulas for two-point
resistances $R_{\alpha\beta^{(i)}}, i=1,2,3$, in terms of
intersection numbers.

It should be noticed that, for two arbitrary nodes $\alpha$ and
$\beta$ such that $\beta\in R_1(\alpha)$, we have
$P_1(x)=\frac{x}{\sqrt{\kappa}}$. Therefore, by using (\ref{l2})
and (\ref{m=1}) we obtain
\begin{equation}\label{alfa1}
L^{-1}_{\beta\alpha}=\frac{r}{\kappa}\sum_{i,i\neq0}\frac{A_ix_i}{\kappa-x_i}=\frac{-r}{\kappa}\sum_{i,i\neq0}A_i+r\sum_{i,i\neq0}\frac{A_i}{\kappa-x_i}.
\end{equation}
Therefore, by using (\ref{eq.res.dist.}), we obtain the following
simple result for all $\beta\in R_1(\alpha)$
\begin{equation}\label{r1}
R_{\alpha\beta^{(1)}}=\frac{2r}{\kappa}\sum_{i,i\neq0}A_i=\frac{2r}{\kappa}(1-A_0)=\frac{2r}{\kappa}(1-\frac{1}{v}),\;\;\
\end{equation}
where, $v$ is the number of vertices of the graph, and in the last
equality we have used the fact that for regular graphs, we have
\begin{equation}
A_0=\frac{1}{v}.
\end{equation}
In the following, we give analytical formulas for calculating
two-point resistances $R_{\alpha\beta^{(2)}}$ and
$R_{\alpha\beta^{(3)}}$, where
$R_{\alpha\beta^{(2)}}(R_{\alpha\beta^{(3)}})$ denotes the mutual
resistances between $\alpha$ and all $\beta\in R_2(\alpha)$ (all
$\beta\in R_3(\alpha))$.

By using (\ref{op}) and that
$P_k=\frac{Q_k}{\sqrt{\omega_1...\omega_k}}$, we have
$P_2(x)=\frac{1}{\sqrt{\omega_1\omega_2}}(x^2-\alpha_1x-\omega_1)$.
Then, from (\ref{l2}) after some simplifications we obtain for
$\beta\in R_2(\alpha)$
\begin{equation}\label{st2}
L^{-1}_{\alpha\beta^{(2)}}=\frac{r}{\sqrt{\omega_1\omega_2\kappa_2}}(-\sum_{i,i\neq0}A_ix_i+(\alpha_1-\kappa)\sum_{i,i\neq0}A_i+\kappa(\kappa-\alpha_1-1)\sum_{i,i\neq0}\frac{A_i}{\kappa-x_i}).
\end{equation}
By substituting $\alpha_1=\kappa-b_1-c_1$ in
$\kappa(\kappa-\alpha_1-1)$, we obtain
\begin{equation}
\kappa(\kappa-\alpha_1-1)=\kappa(b_1+c_1-1)=\kappa b_1.
\end{equation}
Then, the coefficient of the term
$\sum_{i,i\neq0}\frac{A_i}{\kappa-x_i}$ in (\ref{st2}) is
\begin{equation}
\frac{r\kappa b_1}{\sqrt{\omega_1\omega_2\kappa_2}}=\frac{r\kappa
b_1}{\sqrt{\kappa b_1c_2\kappa_2}}=r\sqrt{\frac{\kappa
b_1}{c_2\kappa_2}}=r.
\end{equation}
Therefore, (\ref{st2}) can be written as
\begin{equation}\label{st22}
L^{-1}_{\alpha\beta^{(2)}}=\frac{r}{\sqrt{\omega_1\omega_2\kappa_2}}(-\sum_{i,i\neq0}A_ix_i+(\alpha_1-\kappa)\sum_{i,i\neq0}A_i)+r\sum_{i,i\neq0}\frac{A_i}{\kappa-x_i},
\end{equation}
where, the sum $\sum_{i,i\neq0}A_ix_i$ can be calculated by using
(\ref{st1111}). It can be easily shown that
\begin{equation}\label{Aixii}
\lim_{x\rightarrow\infty}[x^{2}\frac{\partial}{\partial\xi}(\frac{1}{\xi}G_{\mu}(x/\xi))]=a_{d-2}-b_{d-1},
\end{equation}
where, $a_{d-2}$ and $b_{d-1}$ are the coefficients of $x^{d-2}$
and $x^{d-1}$ in $Q^{(1)}_{d-1}$ and $Q_{d}$, respectively. From
the recursion relations (\ref{op}) and (\ref{oq}), one can see
that $a_{d-2}=b_{d-1}=-(\alpha_1+...+\alpha_d)$. Therefore, from
(\ref{st1111}) and (\ref{Aixii}) we obtain
\begin{equation}\label{Aixi1}
\sum_{i,i\neq0}A_ix_i=-A_0\kappa=-\frac{\kappa}{v}.
\end{equation}
Then, by using (\ref{eq.res.}) and (\ref{st22}), one can write
$R_{\alpha\beta^{(2)}}$ as follows
\begin{equation}\label{r2}
R_{\alpha\beta^{(2)}}=\frac{2r}{\sqrt{\omega_1\omega_2\kappa_2}}\{(\kappa-\alpha_1)-\frac{2\kappa-\alpha_1}{v}\},
\end{equation}
where, by using (\ref{intersec}) and (\ref{omegal}), we obtain the
following main result in terms of the intersection numbers of the
graph
\begin{equation}\label{r22}
R_{\alpha\beta^{(2)}}=\frac{2r}{b_0b_1}\{b_1+1-\frac{b_0+b_1+1}{v}\}.
\end{equation}

 Now, consider $\beta\in R_3(\alpha)$. Then,
by using (\ref{op}) and $P_k=\frac{Q_k}{\beta_1...\beta_k}$ we
obtain
$P_3(x)=\frac{1}{\sqrt{\omega_1\omega_2\omega_3}}(x^3-(\alpha_1+\alpha_2)x^2-(\omega_1+\omega_2-\alpha_1\alpha_2)x+\alpha_2\omega_1)$.
As above, after some calculations, we obtain for $\beta\in
R_3(\alpha)$
$$
L^{-1}_{\alpha\beta^{(3)}}=\frac{r}{\sqrt{\omega_1\omega_2\omega_3\kappa_3}}\{\frac{\kappa^2}{v}-2(a_{d-3}-b_{d-2}+b^2_{d-1}-b_{d-1}a_{d-2})-(\alpha_1+\alpha_2-\kappa)\frac{\kappa}{v}-$$
\begin{equation}\label{st3}
(\kappa^2-\kappa(\alpha_1+\alpha_2)-\omega_1-\omega_2+\alpha_1\alpha_2)(\frac{v-1}{v})+
(\kappa^3-\kappa^2(\alpha_1+\alpha_2)-\kappa(\omega_1+\omega_2-\alpha_1\alpha_2)+\alpha_2\omega_1)\sum_{i,i\neq0}\frac{A_i}{\kappa-x_i}\}.
\end{equation}
Again, by substituting $\alpha_1$, $\alpha_2$, $\omega_1$ and
$\omega_2$ from  (\ref{omegal}), we have
\begin{equation}
\frac{1}{\sqrt{\omega_1\omega_2\omega_3\kappa_3}}(\kappa^3-\kappa^2(\alpha_1+\alpha_2)-\kappa(\omega_1+\omega_2-\alpha_1\alpha_2)+\alpha_2\omega_1)=\frac{\kappa
b_1b_2}{\sqrt{\kappa b_1c_2b_2c_3\kappa_3}}=1.
\end{equation}
 Therefore,
(\ref{st3}) can be written as follows
$$
L^{-1}_{\alpha\beta^{(3)}}=\frac{r}{\sqrt{\omega_1\omega_2\omega_3\kappa_3}}\{\frac{\kappa^2}{v}-2(a_{d-3}-b_{d-2}+b^2_{d-1}-b_{d-1}a_{d-2})-(\alpha_1+\alpha_2-\kappa)\frac{\kappa}{v}-$$
\begin{equation}\label{st33}
(\kappa^2-\kappa(\alpha_1+\alpha_2)-\omega_1-\omega_2+\alpha_1\alpha_2)(\frac{v-1}{v})\}+
r\sum_{i,i\neq0}\frac{A_i}{\kappa-x_i}.
\end{equation}

In (\ref{st3}), we have used the following equality
\begin{equation}\label{Aixi}
\lim_{x^2\rightarrow\infty}[x^{3}\frac{\partial^2}{\partial\xi^2}(\frac{1}{\xi}G_{\mu}(x/\xi))]=2(a_{d-3}-b_{d-2}+b^2_{d-1}-b_{d-1}a_{d-2}),
\end{equation}
where, $a_{d-3}$ and $b_{d-2}$ are the coefficients of $x^{d-3}$
and $x^{d-2}$ in $Q^{(1)}_{d-1}$ and $Q_{d}$, respectively. From
the recursion relations (\ref{op}) and (\ref{oq}), one can see
that
$a_{d-3}=\prod_{i<j=1}^d\alpha_i\alpha_j-(\omega_1+...+\omega_{d-1})$
and $b_{d-2}=a_{d-3}-\omega_d$. Therefore, we have
$a_{d-3}-b_{d-2}+b^2_{d-1}-b_{d-1}a_{d-2}=\omega_d$.

Then, by using (\ref{eq.res.}), $R_{\alpha\beta^{(3)}}$ is given
by
\begin{equation}\label{r3s}
R_{\alpha\beta^{(3)}}=\frac{2r}{\sqrt{\omega_1\omega_2\omega_3\kappa_3}}\{\omega_d+(\alpha_1+\alpha_2-2\kappa)\frac{\kappa}{v}+
(\kappa^2-\kappa(\alpha_1+\alpha_2)-\omega_1-\omega_2+\alpha_1\alpha_2)(\frac{v-1}{v})\}.
\end{equation}
In terms of the intersection numbers of the graph, we obtain the
following main result
\begin{equation}\label{r33s}
R_{\alpha\beta^{(3)}}=\frac{2r}{b_0b_1b_2}\{b_{d-1}c_d+
b_2-b_0+c_2+b_1b_2-\frac{(b_0+1)(b_2+c_2)+b_1(b_0+b_2)}{v}\}.
\end{equation}
\subsection{Two-point resistances in infinite regular networks} As
the results (\ref{eq.res.dist.}) and (\ref{st1111}) show, the
two-point resistances on a network depend only on the Stieltjes
function $G_{\mu}(x)$ corresponding to the network. Clearly, each
Stieltjes function  corresponding to an infinite network has a
unique representation as an infinite continued fraction as follows
\begin{equation}\label{sti0}
G_{\mu}(x)=\int_{R}\frac{\mu(dy)}{x-y}=\frac{1}{z-\alpha_0-\frac{\beta_1^2}{z-\alpha_1-\frac{\beta_2^2}
{z-\alpha_2-\frac{\beta_3^2}{z-\alpha_3-\cdots}}}} ,
\end{equation}
where, the sequence $\alpha_0,\alpha_1,...;\beta_1,\beta_2,...$
never stops. It is well known that all infinite continued fraction
expansions as in (\ref{sti0}) can be approximated with finite ones.
Therefore, the two-point resistances on an infinite-size resistor
network can be approximated with those of the corresponding
finite-size networks.

In the following section, we give some examples of finite resistor
networks such that at the limit of the large size of the networks,
we obtain some infinite regular networks. Then, we approximate the
infinite networks with finite ones as illustrated above.

\section{Examples}
In this section we calculate two-point resistances
$R_{\alpha\beta^{(i)}}$, for $i=1,2,3$ by using (\ref{r1}),
(\ref{r22}) and (\ref{r33s}), in some important distance-regular
networks with diameters $d=1$, $d=2$ and $d>2$, respectively..

\subsection{Complete network $K_n$}
The complete network $K_n$ is the simplest example of
distance-regular networks. This network has $n$ vertices with
$n(n-1)/2$ edges, the degree of each vertex is $\kappa=n-1$ also
the network has diameter $d=1$. The intersection array of the
network is $\{c_0;b_1\}=\{n-1;1\}$. Clearly, this graph has only
two strata $\Gamma_0(\alpha)=\alpha$ and
$R_1(\alpha)=\{\beta:\beta\neq\alpha\}$. Then, the only two-point
resistance is $R_{\alpha\beta^{(1)}}$ which is given by using
(\ref{r1}) as follows
\begin{equation}
R_{\alpha\beta^{(1)}}=\frac{2r}{v-1}(1-\frac{1}{v})=\frac{2r}{v}\;\,
\;\;\ \mbox{for all}\;\ \beta\in \Gamma_1(\alpha).
\end{equation}
\subsection{Strongly regular  networks}
One of the most important distance regular  networks are those
with diameter $d=2$, called strongly regular  networks. A  network
with $v$ vertices is strongly regular with parameters $v, \kappa,
\lambda, \mu$ whenever it is not complete or edgeless and \\(i)
each vertex is adjacent to $\kappa$
vertices,\\
 (ii) for each pair of adjacent vertices there are $\lambda$
vertices adjacent to both,\\ (iii) for each pair of non-adjacent
vertices there are $\mu$ vertices adjacent to both.\\
 For a strongly regular network, the intersection
array is given by
\begin{equation}\label{stron0}
\{c_0,c_{1};b_1,b_2\}=\{\kappa,\kappa-\lambda-1;1,\mu\}.
\end{equation}

One can notice that, if we consider  networks with diameter two
and maximum degree $\kappa$ and $\alpha\in V$, then $\alpha$ has
at most $\kappa$ neighbors, and at most $\kappa(\kappa-1)$
vertices lie at distance two from $\alpha$. Therefore
\begin{equation}\label{ineq}
v\leq 1+\kappa+\kappa^2-\kappa=\kappa^2+1,\;\;\ \mbox{or}\;\;\
\kappa\geq\sqrt{v-1},
\end{equation}
where, in the following by using the  inequality (\ref{ineq}), we
will obtain upper bounds for two-point resistances in strongly
regular  networks. To do so, first we calculate two-point
resistances for these  networks.

By using (\ref{r1}), (\ref{r22}) and (\ref{stron0}), we obtain
\begin{equation}\label{str}
R_{\alpha\beta^{(1)}}=\frac{2r}{\kappa}(\frac{v-1}{v}),\;\
\mbox{and}\;\
\end{equation}
\begin{equation}\label{strong}
R_{\alpha\beta^{(2)}}=\frac{2r}{\kappa(\kappa-\lambda-1)}(\kappa-\lambda-\frac{2\kappa-\lambda}{v}),
\end{equation}
respectively. Then, from (\ref{ineq}) and (\ref{str}), we obtain
the following upper bound for $R_{\alpha\beta^{(1)}}$
\begin{equation}\label{upper}
R_{\alpha\beta^{(1)}}\leq \frac{2r\sqrt{v-1}}{v}.
\end{equation}

 Now, we consider the following two well-known
strongly regular
 networks.\\
\textbf{A. Petersen  network}\\
Petersen  network \cite{Ass.sch.} is a strongly regular network
with parameters $(v, \kappa, \lambda, \eta)=(10,3,0,1)$ and the
intersection array $\{c_0,c_{1};b_1,b_2\}=\{3,2;1,1\}$. Therefore,
by using (\ref{str}) and (\ref{strong}), we obtain
\begin{equation}\label{pet}
R_{\alpha\beta^{(1)}}=\frac{3r}{5}\;\ \mbox{and}\;\;\;\
R_{\alpha\beta^{(2)}}=\frac{4r}{5}.
\end{equation}
From (\ref{pet}), it is seen that $R_{\alpha\beta^{(1)}}$ in
Petersen graph saturates the upper bound (\ref{upper}).\\
\textbf{B. Normal subgroup scheme}\\
\textbf{Definition 2.3} The partition $P=\{P_0,P_1,...,P_d \}$ of
a finite group $G$ is called a blueprint \cite{Ass.sch.} if
\\(i) $P_0=\{e\}$\\(ii) for i=1,2,...,d if $g\in P_i$ then $g^{-1}\in
P_i$\\(iii) the set of relations $R_i=\{(\alpha,\beta)\in G\otimes
G|\alpha^{-1}\beta\in P_i\}$ on $G$ form an association scheme.
The set of real conjugacy classes given in Appendix A of Ref.
\cite{js} is an example of blueprint on $G$. Also, one can show
that in the regular representation, the class sums $\bar{P_i}$ for
$i=0,1,...,d$ defined  as
\begin{equation}
\bar{P_i} = \sum_{\gamma\in P_i}\gamma \in CG, \;\;\ i=0,1,...,d,
\end{equation}
are the adjacency matrices of a blueprint scheme.

In \cite{js}, it has been shown that, if $H$ be a normal subgroup
of $G$, the following blueprint classes
\begin{equation}\label{xxx}
P_0=\{e\}, \;\;\;\ P_1=G-\{H\}, \;\;\;\ P_2=H -\{e\},
\end{equation}
define a strongly regular network with parameters $(v, \kappa,
\lambda, \eta)=(g,g-h,g-2h,g-h)$ and the following intersection
array
\begin{equation}
\{c_0,c_{1};b_1,b_2\}=\{g-h,h-1; 1,g-h\},
\end{equation}
 where, $g:=|G|$ and $h:=|H|$. It is interesting to note that in normal subgroup scheme, the
intersections numbers and other parameters depend only on the
cardinalities of the group and its normal subgroup. By using
(\ref{omegal}), the QD parameters are given by
$\{\alpha_1,\alpha_2;\omega_1,\omega_2\}=\{g-2h,0;g-h,(g-h)(h-1)\}$.
Also, from (\ref{xxx}), it is seen that $|\Gamma_2(\alpha)|=h-1$.
Then, by using (\ref{str}) and (\ref{strong}), we obtain
\begin{equation}\label{NS1}
R_{\alpha\beta^{(1)}}=\frac{2r(g-1)}{g(g-h)},\;\ \mbox{and}\;\;\;\
R_{\alpha\beta^{(2)}}=\frac{2r}{(g-h)}.
\end{equation}

One should notice that, the maximum degree $\kappa$  for normal
subgroup scheme is $\kappa_{max}=g-2$ ($h=2$), which can be appear
in networks with even cardinality such as dihedral group.
Therefore, for normal subgroup scheme (strongly regular networks
with parameters $(g,g-h,g-2h,g-h)$), we have
\begin{equation}\label{NSU}
\kappa\leq g-2,
\end{equation}
and therefore, by using (\ref{ineq}), (\ref{NS1}) and (\ref{NSU})
($\kappa=g-h$), we obtain upper and lower bounds for
$R_{\alpha\beta^{(1)}}$ and $R_{\alpha\beta^{(2)}}$ as follows
\begin{equation}\label{NSU}
\frac{2r(g-1)}{g(g-2)}\leq R_{\alpha\beta^{(1)}}\leq
\frac{2r\sqrt{g-1}}{g} \;\;\;\ , \;\;\;\;\;\;\;\;\
\frac{2r}{g-2}\leq R_{\alpha\beta^{(2)}}\leq
\frac{2r}{\sqrt{g-1}}.
\end{equation}

 As an example, we consider the dihedral group $G=D_{2m}$, where its normal subgroup
is $H=Z_m$. Therefore, the blueprint classes are given by
\begin{equation}
P_0=\{e\}, \;\;\ P_1=\{b,ab,a^2b,...,a^{m-1}b\}, \;\;\
P_2=\{a,a^{2},...,a^{(m-1}\},
\end{equation}
which form  a strongly regular network with parameters
$(2m,m,0,m)$ and the following intersection array
\begin{equation}
\{c_0,c_{1};b_1,b_2\}=\{m,m-1;1,m\}.
\end{equation}
By using (\ref{NS1}), we obtain
\begin{equation}\label{NS1}
R_{\alpha\beta^{(1)}}=\frac{r(2m-1)}{m^2},\;\ \mbox{and}\;\;\;\
R_{\alpha\beta^{(2)}}=\frac{2r}{m}.
\end{equation}
\subsection{Cycle network $C_v$}
A cycle network or cycle is a network that consists of some number
of vertices connected in a closed chain. The cycle network with
$v$ vertices is denoted by $C_v$ with $\kappa=2$. For odd
$v=2m+1$, the intersection array is given by
\begin{equation}
\{c_0,...,c_{m};b_1,...,b_m\}=\{2,1,...,1;1,...,1,1\},
\end{equation}
where, for even $v=2m$, we have
\begin{equation}
\{c_0,...,c_{m};b_1,...,b_m\}=\{2,1,...,1,2;1,...,1,2\}
\end{equation}
and the network consists of $m+1$ strata. We consider $v=2m$ (the
case $v=2m+1$ can be considered similarly). Then, by using the
recursion relations (\ref{op}) and (\ref{oq}) one can obtain the
following closed form for the Stieltjes function
\begin{equation}\label{sticycle}
G_{\mu}(x)=\frac{1}{m}\frac{T'_m(x/2)}{T_m(x/2)},
\end{equation}
where, $T_k(x)$ are Tchebyshev polynomials of the first kind.

By using (\ref{r1}), (\ref{r22}) and (\ref{r33s}) we obtain
\begin{equation}\label{cycle}
R_{\alpha\beta^{(1)}}=r(\frac{2m-1}{2m}),\;\;\
R_{\alpha\beta^{(2)}}=2r(\frac{m-1}{m}),\;\ \mbox{and}\;\;\
R_{\alpha\beta^{(3)}}=3r(\frac{2m-3}{2m}),
\end{equation}
respectively. From (\ref{cycle}), one can easily deduce that
\begin{equation}\label{cycle1}
R_{\alpha\beta^{(k)}}=kr(\frac{2m-k}{2m})\;\, \;\;\ k=1,2,...,m.
\end{equation}

At the limit of the large $m$, the cycle network tend to the
infinite line network and the Stieltjes function (\ref{sticycle})
reads as
\begin{equation}
G_{\mu}(x)=\frac{1}{\sqrt{x^2-4}}.
\end{equation}
Then, for $A_0$ we have
\begin{equation}
A_0=\lim_{x\rightarrow2}((x-2)G_{\mu}(x))=0.
\end{equation}
Therefore, by using (\ref{r1}), (\ref{r22}) and (\ref{r33s}) we
obtain
\begin{equation}
R_{\alpha\beta^{(1)}}=r,\;\;\ R_{\alpha\beta^{(2)}}=2r,\;\;\
R_{\alpha\beta^{(3)}}=3r.
\end{equation}
In fact, it can be easily shown that
\begin{equation}
R_{\alpha\beta^{(k)}}=kr,\;\;\ k=1,2,...
\end{equation}
where, this result could be obtained from (\ref{cycle1}), for large
$m$.

As (\ref{cycle1}) indicates, for $m$ larger than $\sim 70$ the
difference $|R_{\alpha\beta^{(k)}}-kr|$ is equal to $\sim 0.01$,
where for $m$ larger than $\sim 100$ we have
$|R_{\alpha\beta^{(k)}}-kr|\sim 0$. Therefore, the finite resistor
network $C_{2m}$, with $m\sim 100$ is a good approximation for the
infinite line resistor network.

\subsection{$d$-cube}
The $d$-cube, i.e. the hypercube of dimension $d$, also called
Hamming cubes, is a network with $2^d$ nodes, each of which can be
labeled by an $d$-bit binary string. Two nodes on the hypercube
described by bitstrings $\vec{x}$ and $\vec{y}$ are are connected by
an edge if $|\vec{x}- \vec{y}|=1$, where $|\vec{x}|$ is the Hamming
weight of $\vec{x}$. In other words, if $\vec{x}$ and $\vec{y}$
differ by only a single bit flip, then the two corresponding nodes
on the graph are connected. Thus, each of the $2^d$ nodes on the
$d$-cube has degree $d$. For the $d$-cube we have $d+1$ strata with
\begin{equation}\label{cubevalenc}
\kappa_i=\frac{d!}{i!(d-i)!}\;\ , \;\ 0\leq i\leq d-1.
\end{equation}
The intersection numbers are given by
\begin{equation}
b_i=d-i,\;\;\ 0\leq i\leq d-1; \;\;\;\ c_i=i,\;\;\ 1\leq i\leq d.
\end{equation}
Then, by using (\ref{r1}), (\ref{r22}) and (\ref{r33s}), we obtain
$$R_{\alpha\beta^{(1)}}=\frac{2^d-1}{d2^{d-1}}r,\;\;\ R_{\alpha\beta^{(2)}}=\frac{2^{d-1}-1}{(d-1)2^{d-2}}r,\;\;\ \mbox{and}$$
\begin{equation}\label{result}
R_{\alpha\beta^{(3)}}=\frac{r}{d(d-1)(d-2)}\{\frac{2^{d}(d^2-2d+2)-3d(d-1)-2}{2^{d-1}}\}.
\end{equation}

 From (\ref{result}) one can see that, at the limit of the large dimension $d$, the
 two-point resistances $R_{\alpha\beta^{(i)}}$, $i=1,2,3$ tend to
 zero. Since $R_{\alpha\beta^{(1)}}$ tend to zero for $d$ larger than $\sim
 200$. Therefore, the finite $d$-cube with $d$ larger than $\sim
 200$ is a good approximation for the infinite hypercube
 resistor network.
\subsection{Johnson network}
Let $n\geq 2$ and $d\leq n/2$. The Johnson network $J(n,d)$ has
all $d$-element subsets of $\{1,2,...,n\}$ such that two
$d$-element subsets are adjacent if their intersection has size
$d-1$. Two $d$-element subsets are then at distance $i$ if and
only if they have exactly $d-i$ elements in common. The Johnson
network $J(n,d)$ has $v=\frac{n!}{d!(n-d)!}$ vertices, diameter
$d$ and the valency $\kappa=d(n-d)$. Its intersection array is
given by
\begin{equation}
b_i=(d-i)(n-d-i), \;\;\;\ 0\leq i\leq d-1; \;\;\ c_i=i^2, \;\;\;\
1\leq i\leq d,
\end{equation}
Then, by using (\ref{r1}), (\ref{r22}) and (\ref{r33s}), we obtain
$$R_{\alpha\beta^{(1)}}=\frac{2(n!-d!(n-d)!)}{d(n-d)n!}r,$$
$$R_{\alpha\beta^{(2)}}=\frac{2r}{d(d-1)(n-d)(n-d-1)}\{d(n-d)-(n-2)+\frac{d!(n-d)!(n-2-2d(n-d))}{n!}\}\;\;\ and$$
$$R_{\alpha\beta^{(3)}}=\frac{2r}{d(d-1)(d-2)(n-d)(n-d-1)(n-d-2)}\{d^2(n-2d+1)+$$
$$(3n-2d(n-d)-10)\frac{d(n-d)d!(n-d)!}{n!}+[d^2(n-d)^2-d(n-d)(3n-9)-4(d-1)(n-d-1)+$$
\begin{equation}\label{resultJ}
2(n-2)(n-4)](1-\frac{d!(n-d)!}{n!})\}.
\end{equation}

It could be noticed that for a give $d$, the result (\ref{resultJ})
show that, at the limit of the large dimension $n$, the
 two-point resistances $R_{\alpha\beta^{(i)}}$, $i=1,2,3$ tend to
 zero. Since $R_{\alpha\beta^{(1)}}$ tend to zero for $n$ larger than $\sim
 200$. Therefore, the finite Johnson network $J(n,d)$ with $n$ larger than $\sim
 200$ is a good approximation for the infinite Johnson
 resistor network.
\section{Two-point resistances in more general networks}
Although, we discussed through the paper only the case of
distance-regular networks, but also the method can be used for any
arbitrary regular network. For calculating two-point resistances, we
need only to know the Stieltjes function $G_{\mu}(x)$. For two
arbitrary nodes $\alpha$ and $\beta$ of the network, we choose one
of the nodes, say $\alpha$, as reference vertex. Then, the Stieltjes
function $G_{\mu}(x)$ can be calculated by using the recursion
relations (\ref{op}) and (\ref{oq}), where, as it has be shown in
\cite{jss}, the coefficients $\alpha_i$ and $\beta_i$, for
$i=1,...,d$ in the recursion relations are obtained by applying the
Lanczos algorithm to the adjacency matrix of the network and the
reference vertex $\ket{\alpha}$. In fact, the adjacency matrix of
the network takes a tridiagonal form in the orthonormal basis
$\{\ket{\phi_i}, i=0,1,...,d\}$ produced by Lanczos algorithm and so
we obtain again three term recursion relations as (\ref{op}). But,
in general, the basis produced by Lanczos algorithm do not define a
stratification basis, in the sense that, a vertex ket $\ket{\beta}$
of the network may be appear in more than one of the base vectors
$\ket{\phi _i}$. In these cases, if $d$ is equal to $v$ (the number
of vertices of the network), one can write each vertex ket
$\ket{\beta}$ uniquely as a superposition of the base vectors
$\ket{\phi_i}$ and calculate two-point resistance $R_{\alpha\beta}$
by calculating the entries $\langle \phi_i|L^{-1}|\alpha\rangle$ for
all $i=0,1,...,d$ as illustrated through the paper. In the most
cases, $d$ is less than $v$. In these cases, we need to obtain some
additional orthonormal base vectors $\{\ket{\psi_i},i=1,..,v-d-1\}$
such that the new bases are orthogonal to the subspace spanned by
$\{\ket{\phi_i}, i=0,1,...,d\}$. One can obtain some such additional
base vectors, by choosing a normalized vector orthogonal to the
subspace spanned by $\{\ket{\phi_i}, i=0,1,...,d\}$ as a new
reference state and applying the Lanczos algorithm to the adjacency
matrix of the network and the new reference state. If the number of
the new orthonormal base vectors still be less than $v-d-1$, we
choose another normalized reference state orthogonal to the subspace
spanned by all previous orthonormal bases and apply the Lanczos
algorithm to the adjacency matrix and the new chosen reference
state. By repeating this process until to obtain $v$ orthonormal
basis, one can solve a system of $v$ equations with $v$ unknowns to
write each vertex ket $\beta$ as superposition of the $v$
orthonormal bases.

\section{Conclusion}
The resistance between two arbitrary nodes in a distance-regular
resistor network was obtained in terms of the Stieltjes transform of
the spectral measure or Stieltjes function associated with the
network and its derivatives. It was shown that the resistances
between a node $\alpha$ and all nodes $\beta$ belonging to the same
stratum with respect to the $\alpha$ are the same. Also, explicit
analytical formulaes for two-point resistances $R_{{\alpha\beta}}$
for $\beta$ belonging to the first, second and third stratum with
respect to the $\alpha$ were driven in terms of the size of the
network and the corresponding intersection numbers. In particular,
the two-point resistances in a strongly regular network with
parameters ($v,\kappa,\lambda,\mu$) were given in terms of these
parameters. Moreover, the lower and upper bounds for two-point
resistances in strongly regular networks was discussed. It was
discussed that, the introduced method can be used not only for
distance-regular networks, but also for any arbitrary regular
network by employing the Lanczos algorithm iteratively.
\newpage
 \vspace{1cm}\setcounter{section}{0}
 \setcounter{equation}{0}
 \renewcommand{\theequation}{A-\roman{equation}}
  {\Large{Appendix A}}\\
In this appendix, we give the two-point resistances
$R_{\alpha\beta^{(i)}}$, $i=1,2,3$ for some important
distance-regular networks with $v\leq 70$.\\\\
\begin{tabular}{|c|c|c|c|c|c|c|}
  \hline
  The network&$v$&Intersection array & Ref.& $R_{\alpha\beta^{(1)}}$ & $R_{\alpha\beta^{(2)}}$ & $R_{\alpha\beta^{(3)}}$ \\
  \hline
  Icosahedron & $12$ & $\{5,2,1;1,2,5\}$ & \cite{14} & $\frac{11r}{30}$ & $\frac{7r}{15}$& $\frac{r}{2}$ \\
  L(Petersen) & $15$ & $\{4,2,1;1,1,4\}$ & \cite{14} & $\frac{7r}{15}$ & $\frac{19r}{30}$& $\frac{2r}{3}$ \\
  Pappus, $3$-cover $K_{3,3}$ & $18$ & $\{3,2,2,1;1,1,2,3\}$ & \cite{14} & $\frac{17r}{27}$ & $\frac{8r}{9}$& $\frac{26r}{27}$ \\
  Desargues & $20$ & $\{3,2,2,1,1;1,1,2,2,3\}$ & \cite{14} & $\frac{19r}{30}$ & $\frac{8r}{9}$& $\frac{59r}{60}$ \\
  Dodecahedron & $20$ & $\{3,2,1,1,1;1,1,1,2,3\}$ & \cite{14} & $\frac{19r}{30}$ & $\frac{9r}{10}$& $\frac{16r}{15}$ \\
  $GH(2,1)$ & $21$ & $\{4,2,2;1,1,2\}$ & \cite{14} & $\frac{10r}{21}$ & $\frac{2r}{3}$& $\frac{5r}{7}$ \\
  Klein & $24$ & $\{7,4,1;1,2,7\}$ & \cite{14} & $\frac{23r}{84}$ & $\frac{9r}{28}$& $\frac{r}{3}$ \\
  $GQ(2,4)\setminus$ spread & $27$ & $\{8,6,1;1,3,8\}$ & \cite{14} & $\frac{13r}{54}$ & $\frac{29r}{108}$& $\frac{5r}{18}$ \\
  $H(3,3)$ & $27$ & $\{6,4,2;1,2,3\}$ & \cite{14} & $\frac{26r}{81}$ & $\frac{31r}{81}$& $\frac{11r}{27}$ \\
  coxeter & $28$ & $\{3,2,2,1;1,1,1,2\}$ & \cite{14} & $\frac{9r}{14}$ & $\frac{13r}{14}$& $\frac{73r}{84}$ \\
  Taylor($P(13)$) & $28$ & $\{13,6,1;1,6,13\}$ & \cite{vandam} & $\frac{27r}{182}$ & $\frac{44r}{273}$& $\frac{91r}{546}$ \\
  Tutte's $8$-cage & $30$ & $\{3,2,2,2;1,1,1,3\}$ & \cite{14} & $\frac{29r}{45}$ &$\frac{14r}{15}$& $\frac{139r}{90}$ \\
  Taylor(GQ$(2,2)$) & $32$ & $\{15,8,1;1,8,15\}$ & \cite{vandam} & $\frac{31r}{240}$ & $\frac{11r}{80}$& $\frac{17r}{120}$ \\
  Taylor($T(6)$) & $32$ & $\{15,6,1;1,6,15\}$ & \cite{vandam} & $\frac{31r}{240}$ & $\frac{101r}{720}$& $\frac{13r}{90}$ \\
  $IG(AG(2,4)\setminus pc)$ & $32$ & $\{4,3,3,1;1,1,3,4\}$ &\cite{7} & $\frac{31r}{64}$ & $\frac{5r}{8}$& $\frac{125r}{192}$ \\
  Wells & $32$ & $\{5,4,1,1;1,1,4,5\}$ & \cite{7} & $\frac{31r}{80}$ & $\frac{15r}{32}$& $\frac{39r}{80}$ \\
  Hadamard graph & $32$ & $\{8,7,4,1;1,4,7,8\}$ & \cite{9} & $\frac{31r}{128}$ & $\frac{15r}{56}$& $\frac{249r}{896}$ \\
  Odd($4$) & $35$ & $\{4,2,1;1,1,4\}$ & \cite{3},\cite{16} & $\frac{17r}{35}$ & $\frac{22r}{35}$& $\frac{242r}{315}$ \\
  Sylvester & $36$ & $\{5,4,4;1,1,4\}$ & \cite{3} & $\frac{7r}{18}$ & $\frac{17r}{36}$& $\frac{181r}{240}$ \\
  Taylor($P(17)$) & $36$ & $\{17,8,1;1,8,17\}$ & \cite{vandam} & $\frac{35r}{306}$ & $\frac{149r}{1224}$& $\frac{279r}{272}$ \\
  $3$-cover $K_{6,6}$& $36$ & $\{6,5,4,1;1,2,5,6\}$& \cite{vandam}& $\frac{35r}{108}$& $\frac{17r}{45}$& $\frac{211r}{540}$\\
\end{tabular}\\
  \begin{tabular}{|c|c|c|c|c|c|c|}
  The network & $v$ & Intersection array & Ref. &$R_{\alpha\beta^{(1)}}$ & $R_{\alpha\beta^{(2)}}$ & $R_{\alpha\beta^{(3)}}$ \\
  \hline
  $SRG\setminus$ spread& $40$ & $\{9,6,1;1,2,9\}$ & \cite{10} & $\frac{13r}{60}$ & $\frac{11r}{45}$& $\frac{r}{4}$ \\
  $Ho-Si_2(x)$ & $42$ & $\{16,5,1;1,1,6\}$ & \cite{3} & $\frac{41r}{126}$ & $\frac{8r}{21}$& $\frac{34r}{105}$ \\
  Mathon (Cycl$(13,3)$) & $42$ & $\{13,8,1;1,4,13\}$ & \cite{17} & $\frac{41r}{273}$ & $\frac{89r}{546}$& $\frac{91r}{13}$ \\
  $GO(2,1)$ & $45$ & $\{4,2,2,2;1,1,1,2\}$ & \cite{7} & $\frac{22r}{45}$ & $\frac{32r}{45}$& $\frac{4r}{5}$ \\
  $3$-cover $GQ(2,2)$ & $45$ & $\{6,4,2,1;1,1,4,6\}$ & \cite{7} & $\frac{44r}{135}$ & $\frac{107r}{270}$& $\frac{221r}{540}$ \\
  Hadamard graph & $48$ & $\{12,11,6,1;1,6,11,12\}$ & \cite{vandam} & $\frac{47r}{288}$ & $\frac{23r}{132}$& $\frac{565r}{3168}$ \\
  $IG(AG(2,5)\setminus pc)$ & $50$ & $\{5,4,4,1;1,1,4,5\}$ & \cite{7} & $\frac{49r}{125}$ & $\frac{12r}{25}$& $\frac{123r}{250}$ \\
  Mathon (Cycl$(16,3)$) & $51$ & $\{16,10,1;1,5,16\}$ & \cite{17} & $\frac{25r}{204}$ & $\frac{89r}{680}$& $\frac{68r}{510}$ \\
  $GH(3,1)$& $52$ & $\{6,3,3;1,1,2\}$ & \cite{13} & $\frac{51r}{156}$ & $\frac{11r}{26}$& $\frac{23r}{52}$ \\
  Taylor(SRG$(25,12)$)& $52$ & $\{25,12,1;1,12,25\}$ & \cite{vandam} & $\frac{51r}{650}$ & $\frac{319r}{3900}$& $\frac{13r}{156}$ \\
  $3$-cover $K_{9,9}$&$54$ & $\{9,8,6,1;1,3,8,9\}$ & \cite{vandam} & $\frac{53r}{243}$ & $\frac{13r}{54}$& $\frac{239r}{972}$ \\
  Gosset,Tayl(Schl$\ddot{a}$fli)&$56$ & $\{27,10,1;1,10,27\}$ & \cite{vandam} & $\frac{55r}{756}$ &$\frac{289r}{3780}$& $\frac{49r}{630}$ \\
  Taylor(Co-Schl$\ddot{a}$fli)&$56$ & $\{27,16,1;1,16,27\}$ & \cite{vandam} & $\frac{55r}{756}$ & $\frac{227r}{3024}$& $\frac{11r}{144}$ \\
  Perkel&$57$ & $\{6,5,2;1,1,3\}$& \cite{3},\cite{5} & $\frac{56r}{171}$ & $\frac{22r}{57}$& $\frac{68r}{171}$ \\
  Mathon(Cycl$(11,5)$)&$60$ & $\{11,8,1;1,2,11\}$ &\cite{17} & $\frac{59r}{330}$ & $\frac{32r}{165}$& $\frac{r}{5}$ \\
  Mathon(Cycl$(19,3)$)&$60$ & $\{19,12,1;1,6,19\}$ & \cite{17} & $\frac{59r}{570}$ & $\frac{187r}{1710}$& $\frac{3151r}{570}$ \\
  Taylor($SRG(29,14)$)&$60$ & $\{29,14,1;1,14,29\}$ & \cite{vandam} & $\frac{59r}{870}$ & $\frac{214r}{3045}$& $\frac{29r}{406}$ \\
  $GH(2,2)$ & $63$&$\{6,4,4;1,1,3\}$ & \cite{13} & $\frac{62r}{189}$ & $\frac{76r}{189}$& $\frac{271r}{504}$ \\
  $H(3,4)$,Doob&$64$& $\{9,6,3;1,2,3\}$ & \cite{vandam} & $\frac{7r}{32}$ & $\frac{r}{4}$& $\frac{25r}{96}$ \\
  Locally Petersen&$65$&$\{10,6,4;1,2,5\}$ &  & $\frac{64r}{325}$ & $\frac{73r}{325}$& $\frac{49r}{156}$ \\
  Doro & $68$ &$\{12,10,3;1,3,8\}$&  & $\frac{67r}{408}$ & $\frac{145r}{816}$& $\frac{253r}{1020}$ \\
  Doubled Odd($4$)& $70$&$$\{4,3,3,2,2,1,1;$$
  $$1,1,2,2,3,3,4\}$$& \cite{vandam}& $\frac{69r}{140}$& $\frac{68r}{105}$& $\frac{869r}{1260}$\\
  $J(8,4)$& $70$&$\{16,9,4,1;1,4,9,16\}$ & \cite{vandam} & $\frac{69r}{560}$ & $\frac{337r}{2520}$& $\frac{691r}{5040}$ \\
  \hline
\end{tabular}

\end{document}